\documentclass[a4paper,10pt]{article}
\usepackage{graphicx}
\usepackage{dcolumn}
\usepackage{bm}
\usepackage{amsmath}
\usepackage{amssymb}

\begin{document}
\title{Reflection-less device allows electromagnetic warp drive}
\author{T. Ochiai$^1$ and J.C. Nacher$^2$}   

\maketitle

\begin{center}
{\it $^1$ School of Social Information Studies, Otsuma Women's University }
\end{center}
\begin{center}
{\it 2-7-1 Karakida, Tama-shi, Tokyo 206-8540, Japan}
\end{center}
\begin{center}
ochiai@otsuma.ac.jp
\end{center}
\begin{center}
{\it $^2$ Department of Complex and Intelligent Systems, Future University Hakodate}
\end{center}
\begin{center}
{\it 116-2 Kamedanakano-cho, Hakodate, 041-8655, Hokkaido, Japan}
\end{center}
\begin{center}
nacher@fun.ac.jp
\end{center}

\abstract{\bf
One of the striking properties of artificially structured materials is the negative refraction, an optical feature that known natural materials do not exhibit. Here, we propose a simple design, composed of two parallel layers of materials with different refraction indices $n_1=-n_2$, that constructs perfect reflection-less devices. The electromagnetic waves can tunnel from one layer to the other, a feature that resembles a truncation of the physical space leading to an electromagnetic warp drive. Since the refractive indices do not require any large values, this method demonstrates for the first time the practical feasibility of guiding electromagnetic fields in complete absence of reflection phenomena and without degradation of transmission efficiency at all.}
\newline

\newpage

The recent advancement of meta-materials has extended electromagnetic 
properties and led to promising applications in disparate fields from telecommunications and semi-conductor engineering to medical 
imaging, defense and cloaking devices  \cite{exp2,exp3,meta, pendry2006, exp1, leon1, review, liu, ergin}. Metamaterials show new properties like the 
negative refraction, an optical feature that common materials do not exhibit \cite{vesa}. Veselago showed that a material characterized by negative values of 
effective permittivity and permeability exhibits several reversed 
physics phenomena like the reversal of Snell's law. However, and in spite 
of its physical interest, the application of negative 
refraction index has been mainly confined to the construction of perfect 
lens \cite{pendry2000} and the modeling of perfect cloaking devices 
\cite{our, new}. Here, we propose a simple design that constructs perfect reflection-less devices where electromagnetic waves can tunnel from one side to the other side of the device. This perfect electromagnetic tunneling feature resembles a truncation of the physical space leading to a warp drive. 
The device consists of two parallel layers of materials with different refraction indices. While the first layer has a value $n_1$, the second layer consists of  $n_2=-n_1$. Furthermore, the absolute value of the refraction index can take any value, allowing practical applications using values close to 1. 
This simple device displays a striking reflection-less feature, with electromagnetic tunneling effect, that astonishingly has not been reported before. Since the values for the refractive indices do not require any large value, this method demonstrates for the first time the practical possibility 
to guide electromagnetic fields in complete absence of reflection 
phenomena. While perfect couplers have only been suggested using epsilon-near-zero (ENZ) medium \cite{warp}, here we show that any pair of positive and negative refractive index can guide electromagnetic waves through a material without degradation of transmission efficiency at all.

\begin{figure}[htbp]
 \begin{minipage}{0.5\hsize}
  \begin{center}
   \includegraphics[scale=0.27]{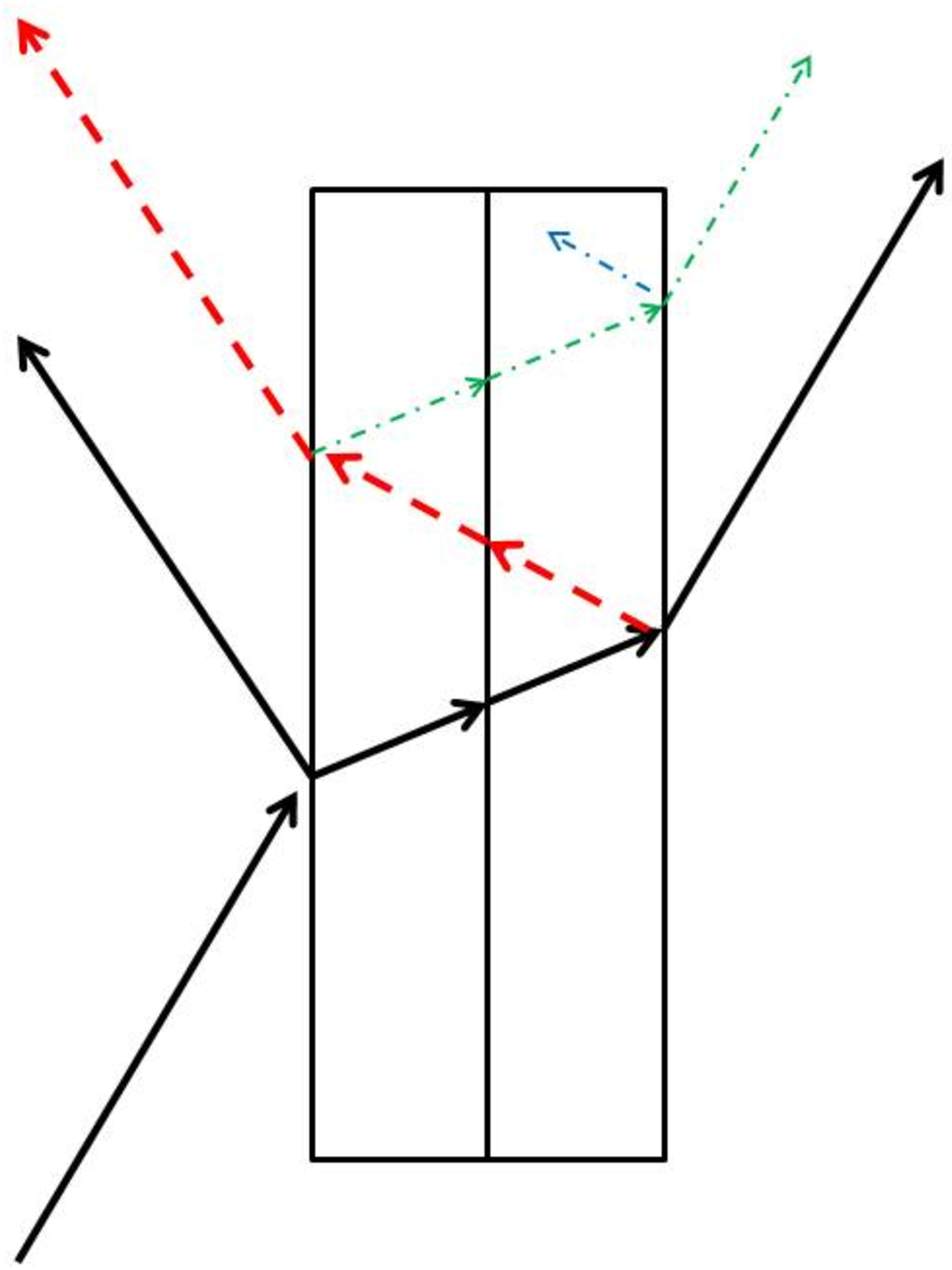}
   \put(-220,160){(a)}
   \put(-150,45){$n_1$}
   \put(-125,45){$n_2$}
   \put(-165,170){$B1$}
   \put(-140,170){$B2$}
   \put(-115,170){$B3$}
  \end{center}
 \end{minipage}
 \begin{minipage}{0.5\hsize}
  \begin{center}
   \includegraphics[scale=0.27]{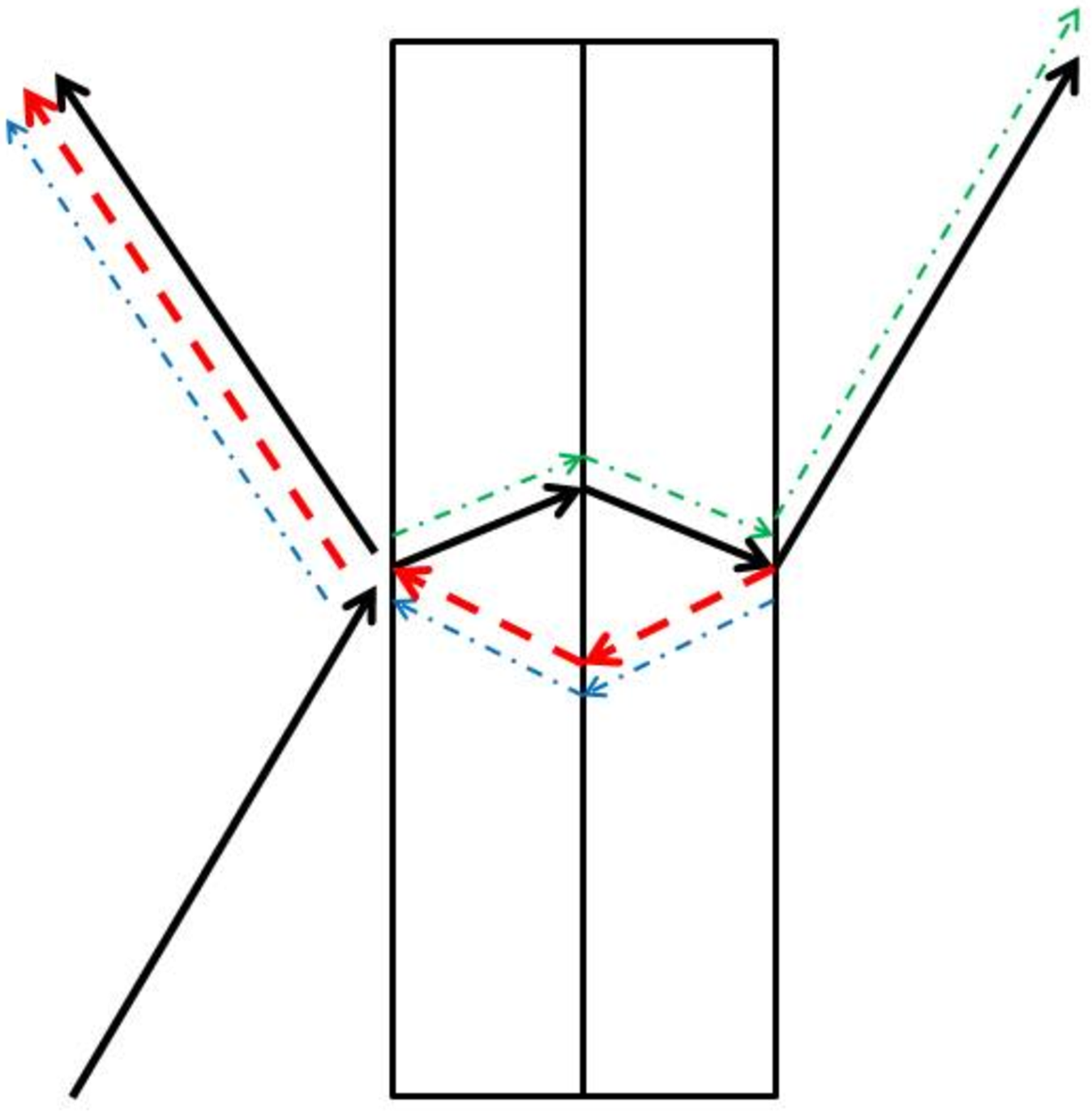}
   \put(-220,160){(b)}
   \put(-150,45){$n_1$}
   \put(-125,45){$n_2$}
   \put(-220,120){\scriptsize perfect}
   \put(-220,110){\scriptsize cancelation}
   \put(-165,170){$B1$}
   \put(-140,170){$B2$}
   \put(-115,170){$B3$}
  \end{center} 
 \end{minipage}
 \caption{(a) The trajectory of light rays through media with refractive index $n_1=n_2=2$.  It represents a typical example of refraction and reflection in conventional materials.  (b) The trajectory of light rays in our proposed device through media with refractive index $n_1=2$ and $n_2=-2$. The line colors correspond with those of (a), but having different light path.  The black solid line represents an incident light ray entering materials $n_1$, $n_2$. The light ray is reflected from B1, bent at B2 and refracted at B3. Then the light ray (red dashed line) is reflected again  from B3 and goes out to left vacuum through B1.  Next, the light ray is reflected again from B1 (green dash-dotted line) and a certain fraction of the light goes out to right vacuum through B3.  The remaining light (blue dash-dotted line) is reflected at B3 again and goes back to the left vacuum through B1. All the reflected and outgoing light rays start at the same point
respectively.}
 \label{fig:image}
\end{figure}

\begin{figure}[htbp]
 \begin{minipage}{0.5\hsize}
  \begin{center}
   \includegraphics[scale=0.4]{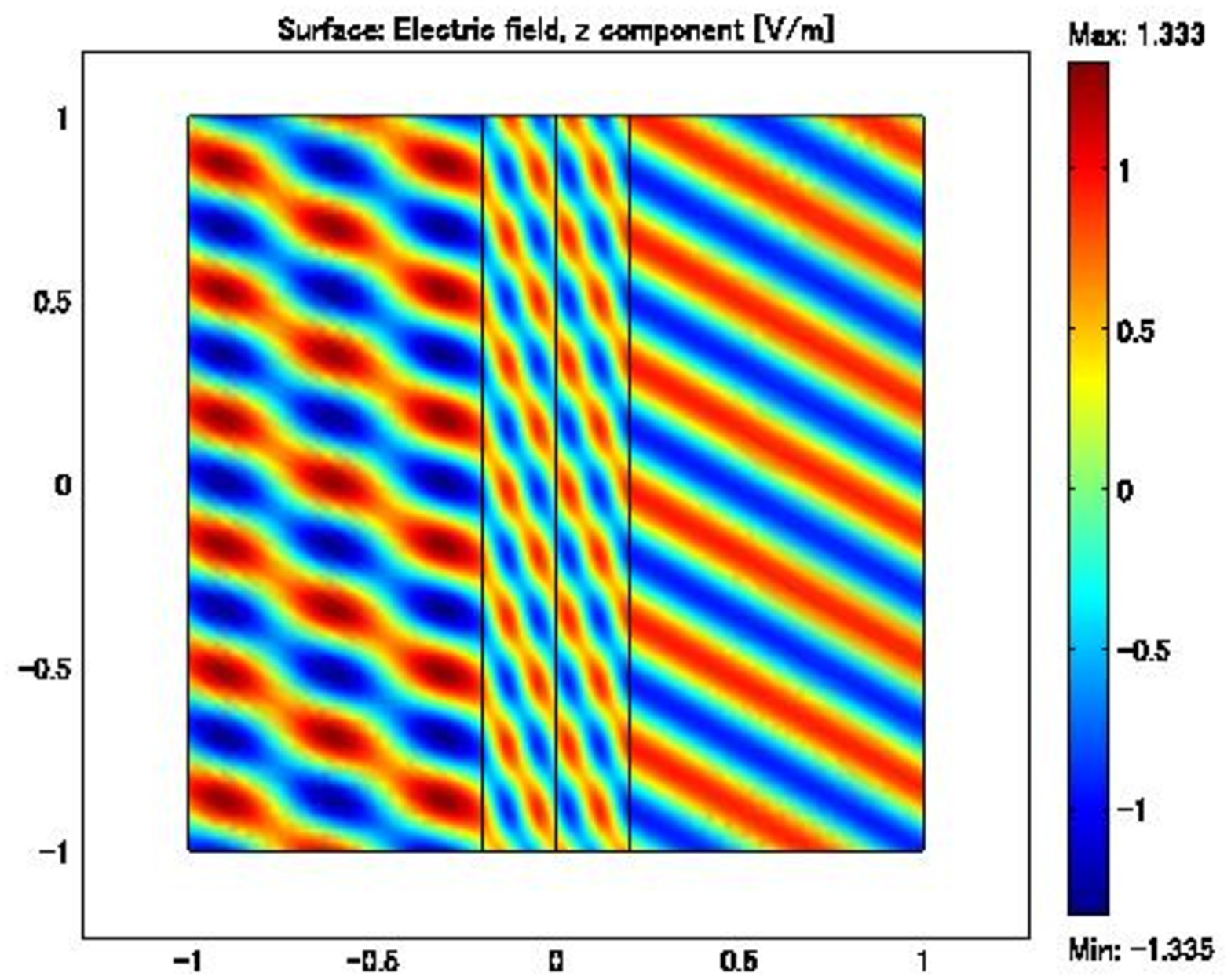}
   \put(-300,240){(a)}
  \end{center}
 \end{minipage}
 \begin{minipage}{0.5\hsize}
  \begin{center}
   \includegraphics[scale=0.4]{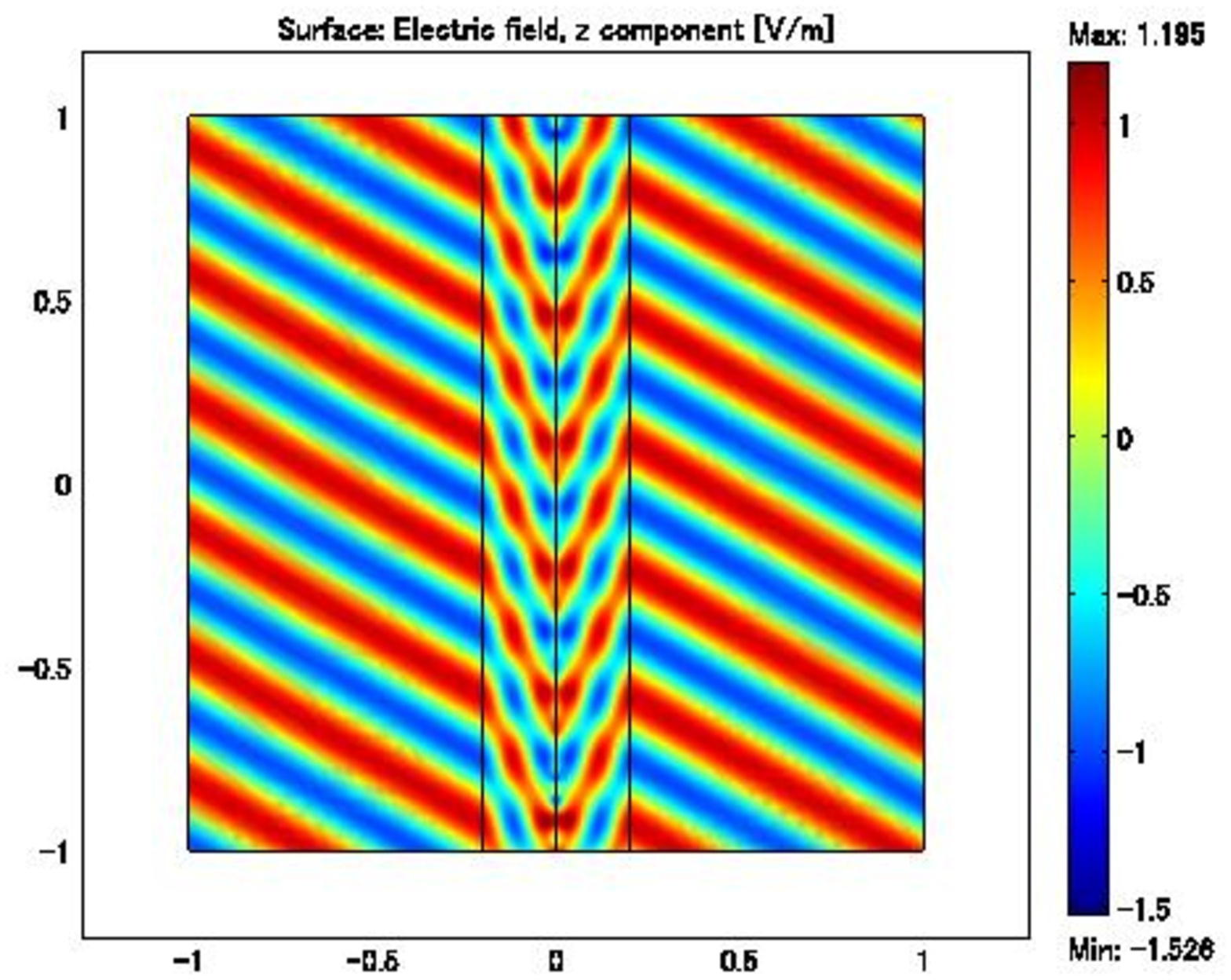}
   \put(-300,240){(b)}
  \end{center} 
 \end{minipage}
\begin{minipage}{0.5\hsize}
  \begin{center}
  \includegraphics[scale=0.4]{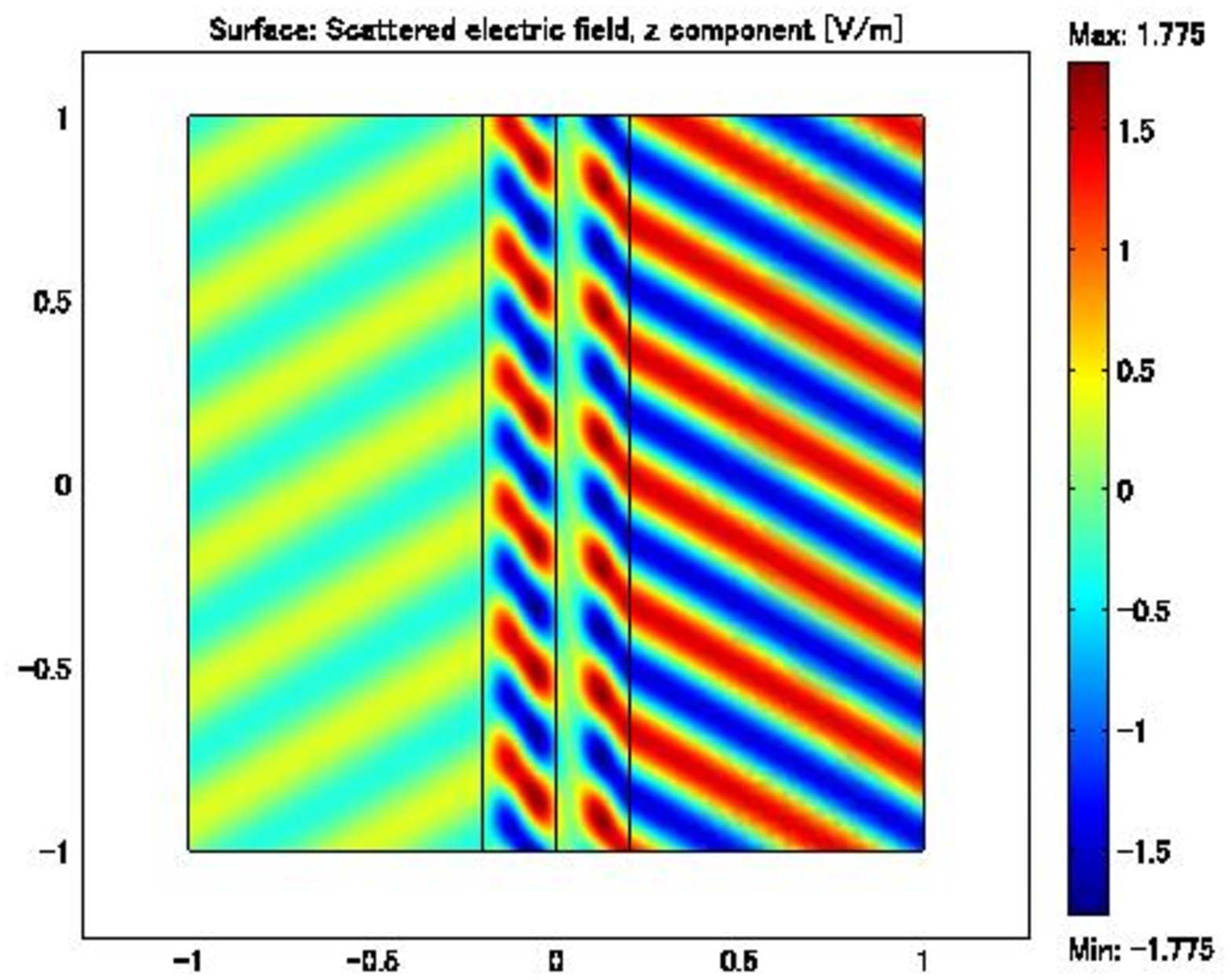} 
  \put(-300,240){(c)}
  \end{center}
 \end{minipage}
 \begin{minipage}{0.5\hsize}
  \begin{center}
   \includegraphics[scale=0.4]{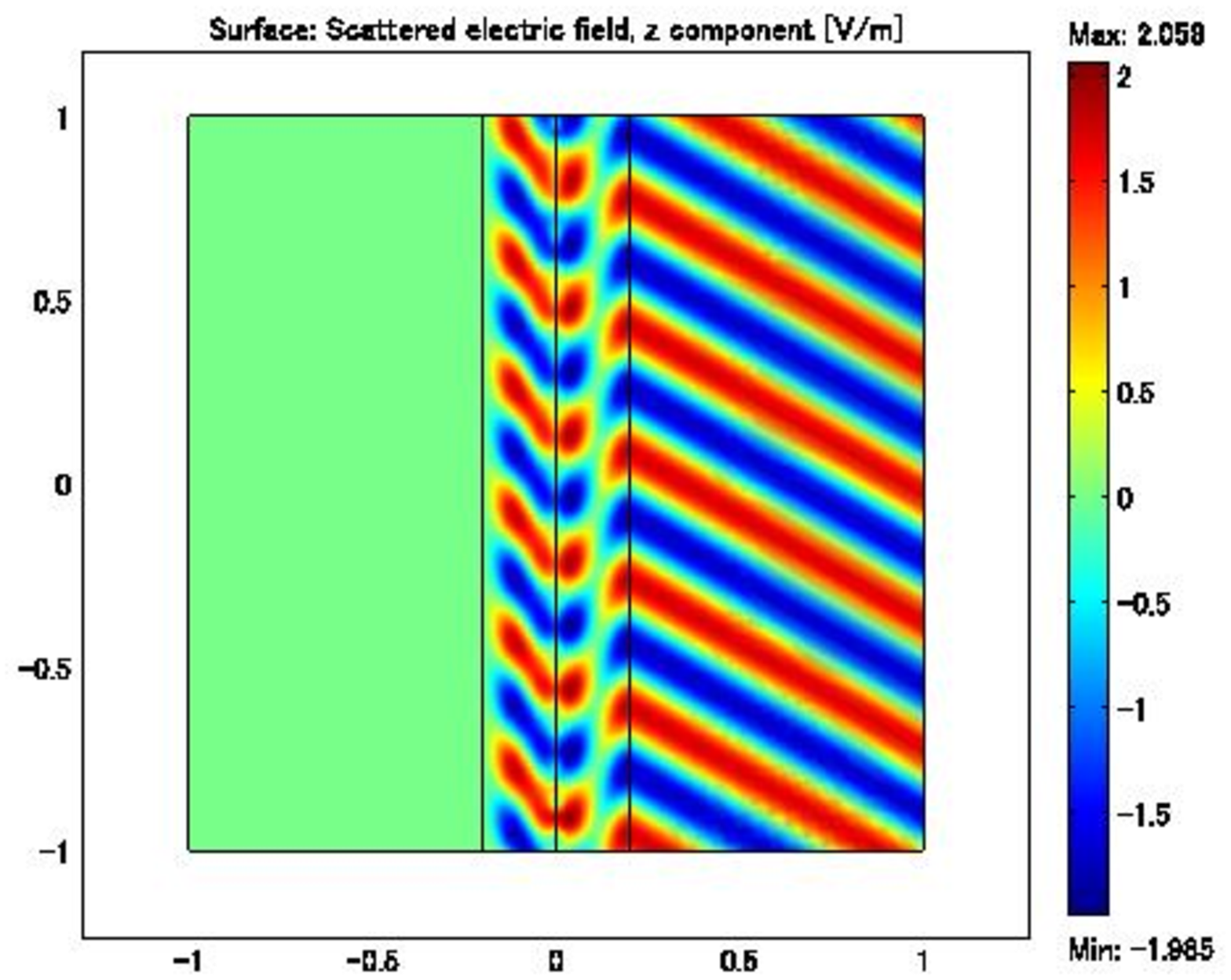}
   \put(-300,240){(d)}
  \end{center}  
 \end{minipage}
 \vspace{-2cm}
 \caption{In all cases, the incoming plane waves, defined as TE mode and with plane of incidence parallel to $xy$-plane, enter a material from left to right with incident angles ($\pi/3$). 
(a) $n_1=2$, $n_2=2$ case. $E_z$-component is  represented as an  incident light wave entering the device from left to right. A wave distortion in left vacuum is observed due to the combination of incident wave and reflected wave. (b) The proposed device with $n_1=2$, $n_2=-2$.  $E_z$-component is represented as in (a), but with opposite sign of $n_2$. The wave pattern of the left vacuum is exactly the same as that of the right vacuum. It shows a perfect tunneling effect or warp drive (i.e., no reflection and perfect transmission of light). (c) $n_1=2$, $n_2=2$ case.  The scattered $E_z$-component of case (a) is represented (i.e., this is the same as (a), but we remove incident wave and show only scattered component). It highlights the existence of reflection in the left vacuum. (d) $n_1=2$, $n_2=-2$. The scattered $E_z$-component of case (b) is represented (i.e., this is the same as (b), but we remove incident wave and show only scattered component). There is no reflected wave at all in the left vacuum.}
 \label{fig:simulation}
\end{figure}

Figure \ref{fig:image}(a) shows a classical reflection phenomena. An electromagnetic wave (light rays) strikes a material and a reflected 
ray is observed. The law of reflection shows that  the angle of incidence equals the angle of 
reflection $\theta_i = \theta_r$.  Physical laws show that reflection of light always occurs when light travels from a material of a given refractive index 
into a medium with a different refractive index. In general, only a fraction of the light is refracted since there is an unavoidable reflection from the 
interface. In sharp contrast, here we show that the proposed device is able to completely suppress the reflected 
light, allowing a perfect transmission of electromagnetic waves from one boundary (B1) to the other boundary (B3) (see Figure \ref{fig:image}(b)). The only requirement is to have two media with different sign of refraction index but with the same absolute value as follows:

\begin{eqnarray}
n(x,y,z)= \left\{
         \begin{array}{c r}
           n_1 & (-a<x<0) \\
	   n_2 & (0<x<a) 
         \end{array}
    \right.
\end{eqnarray}
where $n_1$ and $n_2$ are the refractive indices of the left and right-hand side materials with the condition:
\begin{eqnarray}
n_1=-n_2. \label{eqn:condition}
\end{eqnarray}
Here, we put the device in the uniform media ($n=1$, for example, vacuum or air.) The device has three discontinuous boundaries of refractive index: the boundary between vacuum and the material $n_1$  (i.e., B1 ($x=-a$)), the boundary between $n_1$ and $n_2$  (i.e., B2 ($x=0$)), and the boundary between $n_2$ and vacuum (i.e., B3 ($x=a$)) (see Fig. \ref{fig:image}). It is well known that a boundary between two media having different refractive index causes a reflection, which leads to degrade the transmission 
efficiency (Fig. \ref{fig:simulation} (a)). However, here we show that in spite that our device has three boundaries (B1, B2 and B3), where refraction 
indices change, the resulting reflections are surprisingly and completely canceled each other out. This striking  result is rooted in the combination of plus 
and minus refraction index \cite{our, new}.

Let us consider the general case of the boundary between two different materials 1 and 2. Let $\epsilon_1$ and $\mu_1$ (resp. $\epsilon_2$ and $\mu_2$)
 be permittivity and permeability of material 1 (resp. material 2). Then, the refractive index of material 1 (resp. material 2) 
 is given by $n_1=\sqrt{\epsilon_1\mu_1}$ (resp. $n_2=\sqrt{\epsilon_2\mu_2}$). By considering the standard electromagnetic theory \cite{book3}, for 
the impedance matching case (
$
\sqrt{\frac{\mu_1}{\epsilon_1}}=\sqrt{\frac{\mu_2}{\epsilon_2}}
$)
, the transmission ratio is given by	
\begin{eqnarray}
T(\theta,\phi)=\frac{E_t}{E_i}=\frac{2\cos \theta}{\cos \theta+\cos \phi} \label{eqn:T(i,j)}
\end{eqnarray}
where $\theta$ is the incoming angle of light and $\phi$ is the outgoing (i.e., refraction) angle of light.  Similarly, the reflection ratio is given by
\begin{eqnarray}
R(\theta,\phi)=\frac{E_r}{E_i}
=\frac{\cos \theta-\cos \phi}{\cos \theta+\cos \phi}\label{eqn:R(i,j)}
\end{eqnarray}
 First, we analyze the discontinuity of the boundary B2
 (i.e., the boundary between positive and negative refraction indices). By simply applying Snell's law to Eq. (\ref{eqn:condition}), we obtain $\theta=-\phi$. In this case, by considering Eq. (\ref{eqn:R(i,j)}), we can easily see that there is no reflection at all for all incident angles and polarizations at B2.

Secondly, let us consider that the light enters from vacuum to material $n_1$ through boundary B1 (see continuous black line in Fig. \ref{fig:image}(b)). At a first look at Eq. (\ref{eqn:R(i,j)}), it seems that there should be some reflections at boundary B1. This contribution is given by Eq. (\ref{eqn:R(i,j)}).  However, we will show that this reflection is completely canceled out by summing up all the reflections that take place at the boundary B3.

Next, let us consider the case when the light enters from left vacuum to material $n_1$ through the boundary B1. This light ray is bent at the interface B2 and is reflected from the interface B3. Then, it returns 
to the interface B1 and goes out to left vacuum through the interface B1 (see red dashed line in Fig. \ref{fig:image}(b)). In this case, the ratio is given by
\begin{eqnarray}
&&A_1=T(\theta,\phi)R(\phi,\theta)T(\phi,\theta)
\end{eqnarray}

However, the light rays can be reflected again from the interface B1 and go
and back two times between the interfaces B1 and B3, and finally come back to vacuum through the interface B1 (see green and blue dashed-dotted lines in Fig. \ref{fig:image}(b)). For the case that 
the light rays go and back two times, the ratio is given by $A_2=T(i,j)R^3(j,i)T(j,i)$. In fact, this process leads to an infinite series of contribution to the reflection ratio. Thus, more generally, for going back $n$ times, the ratio can read as 
$A_n=T(i,j)R^{2n-1}(j,i)T(j,i)$. Therefore, the total sum of all contributions is
\begin{eqnarray}
A_{tot}
&=&T(\theta,\phi)\{\sum_{i=1}^{\infty}R^{2n-1}(\phi,\theta)\}T(\phi,\theta)\nonumber\\
&=&\frac{\cos \phi-\cos \theta}{\cos \theta+\cos \phi}
\end{eqnarray}

Unexpectedly, the sum of all contributions cancels the original reflection $R(\theta,\phi)$ of Eq. (\ref{eqn:R(i,j)}). Therefore, there is no 
 reflection at boundary B1 at all.

We can also discuss about the transmission ratio. The leading contribution comes from the case that the light enters from vacuum to the material $n_1$ and goes through $n_1$ and $n_2$ and finally goes out through the interface B3 (see continuous black line in Fig. \ref{fig:image}(b). This contribution of transmission ratio is given by
\begin{eqnarray}
&&B_1=T(\theta,\phi)T(\phi,\theta)
\end{eqnarray}
By following a similar argument, we can illustrate another cases. The light enters into $n_1$ through B1 and is reflected from B3. Then, it goes back to B1 and is reflected again.  Finally, it returns to right vacuum through B3. This contribution is given by $B_2=T(\theta,\phi)R^2(\phi,\theta)T(\phi,\theta)$. Indeed, there are also infinite series of reflections and refractions that contribute to the transmission ratio. The total transmission ratio reads as
\begin{eqnarray}
B_{tot}
&=&T(\theta,\phi)\{\sum_{i=0}^{\infty}R^{2n}(\phi,\theta)\}T(\phi,\theta)\nonumber\\
&=&1
\end{eqnarray}
which shows that the light is completely transmitted from B1 to B2.

On the other hand, it is worth mentioning how the phase velocity behaves in the device. The negative refractive index has opposite sign of phase velocity to the positive refractive index. The phase shift in the 
positive refractive index region $n_1$ is exactly canceled out by the phase shift in the negative refractive index region $n_2$. Therefore, 
in the proposed device, the phase seems to jump instantaneously from one side to another.

In conclusion, there is no reflection for all the interfaces B1, B2 and B3 and phase seems to jump from one side (B1) to the other side (B3) instantaneously.  
The device  behaves as a perfect electromagnetic tunneling device. More concretely, the physical space 
between B1 and B2 seems to be truncated leading to a warp drive.

\begin{figure}
\includegraphics[scale=0.4]{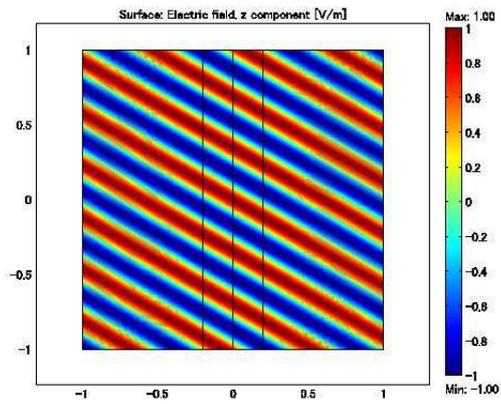}
 \vspace{-2cm}
\caption{\label{fig:vacuum} For comparison with  Fig. \ref{fig:simulation}, $E_z$-component is represented for case $n_1=n_2=1$ (i,e., there is no device. The simulation settings are the same as Fig. \ref{fig:simulation}. }
\end{figure}

In order to demonstrate our theoretical findings, we have conducted a computer simulation using commercial software COMSOL. We simulated 
incident plane waves at a given angle of incidence and measured the $E_z$ component. We set two computational experiments to show the reflection-less effect. The 
first simulation contains two materials with the same refraction index ($n_1=n_2=2$). Incoming waves strike the medium, change 
trajectory according Snell's law and are finally transmitted (see Fig. \ref{fig:simulation}(a)). The subtraction of the incident wave reveals the existence 
of a reflected wave as predicted by the physical laws (see Fig. \ref{fig:simulation}(c)).  It is worth noticing that the $E_z$ component of the left hand side vacuum in Fig. \ref{fig:simulation}(a) is deformed by a combination of incident wave and reflected wave. Now, let us consider a second experiment. Here, although we also 
have two media, the refractive index has the same absolute value but different sign ($n_1=2$ and $n_2=-2$). This difference leads to a striking result. As shown 
in Fig. \ref{fig:simulation}(b), the incident and transmitted waves are identical. This is highlighted in Fig. \ref{fig:simulation}(d), because when an incident wave is removed, no reflected wave is observed in the left vacuum, showing no degradation of transmission efficiency at all. We show the incoming wave 
in vacuum without device in Fig. \ref{fig:vacuum} for comparison purposes.

 It is worth mentioning that our device has also a so-called electromagnetic tunneling effect. Recent works have shown that a material with 
electric permittivity close to zero can behave like a perfect coupler \cite{liu}, where electromagnetic waves can tunnel through a material. This 
effect also leads to an absence of degradation efficiency. However, these proposed theoretical analyses and experiments are based on 
close to zero values for permittivity. In contrast, our device behaves exactly like an electromagnetic tunneling device for any range 
of permittivity values. Figs. \ref{fig:simulation}(b) and \ref{fig:simulation}(d) show that the electromagnetic wave performs a warp drive from one side to another.

Beyond purely electromagnetic devices, reflection also occurs at the surface of transparent media, such as glasses. Many real devices common 
in our daily life have screens and surfaces that cannot escape from reflection phenomena. Several examples are iphone, television and PC screens, 
car glasses and optical devices in general. In particular, for higher incident angles, the reflection is so intense that can make the screen 
unreadable even though we use anti-glare screen protectors. It seems impossible to eliminate the effect of all reflection for any angle and 
any polarization in isotropic media.  Although a design for a reflection-less screen seems impossible, our findings shows that 
a reflection-less isotropic device is possible simply by combining a positive and negative refraction index. Amazingly, at each boundary, there 
is no reflection at all for any arbitrary and polarization angle. 
The structure and design of our proposed design is very simple, and its practical realization is anticipated since it only requires one condition: any 
pair of positive and negative refraction index with the same absolute value. In addition to the mentioned screen and glasses applications for 
visual range of wavelength, the striking cancellation of reflection phenomena could be very fundamental to the design of many novel applications 
in electromagnetic devices from microwave passive devices and defense systems to solar panels and wireless communications wherever 
distortions arising from reflections degrade the transmission efficiency.  
\newline


\noindent{\bf Acknowledgments}
T.O. and J.C.N. gratefully acknowledge the funding support of a Grant-in-Aid for Scientific Research (C) from MEXT, Japan.

\newpage

\end{document}